\documentstyle[epsf,12pt]{article} \newlength{\extraspace}
 \newlength{\extraspaces}
\def\be{\begin{equation}}
\def\ee{\end{equation}}
\def\ba{\begin{eqnarray}}
\def\ea{\end{eqnarray}}
\def\lapx{\,\,\lower 2pt \hbox{$\buildrel<\over{\scriptstyle{\sim}}$}\,\,}
\def\gapx{\,\,\lower 2pt \hbox{$\buildrel>\over{\scriptstyle{\sim}}$}\,\,}

\begin{document} \pagestyle{empty} \begin{titlepage} \begin{flushright}
UTPT-96-12\\ hep-ph/9606338\end{flushright} \vspace{2.5cm} \begin{center}
{\LARGE Hints of Dynamical Symmetry Breaking?}\\\vspace{40pt} {\large B.
Holdom\footnote{holdom@utcc.utoronto.ca}} \vspace{0.5cm}

{\it Department of Physics\\ University of Toronto\\ Toronto,
Ontario\\Canada M5S 1A7} \vspace{0.5cm}

\vskip 2.1cm \vspace{25pt}

\vspace{12pt} \baselineskip=18pt \begin{minipage}{5in}

There is current interest in a possible new massive gauge boson $X$ which mixes
slightly with the $Z$ boson and accounts for certain anomalies in the LEP data.
We show why constraints on models in which the $X$ boson does not couple to the
first two families suggest dynamical electroweak symmetry breaking. The associated
TeV mass fermions make up a fourth family. Constraints on the effects of the fourth
left-handed neutrino also suggest a dynamical origin for its Majorana mass. We
finally comment on related implications for the origin of quark masses.

\end{minipage} \end{center} \vfill \end{titlepage} \pagebreak
\baselineskip=18pt \pagestyle{plain} \setcounter{page}{1}

\section{$Z$--$X$ Mixing}

We have learned by experience that discrepancies between experiment and the
standard model tend to go away over time, and so it is natural to take a cautious
attitude toward the present set of anomalies in the data. But we may still ask,
from a more theoretical point of view, which of the present anomalies are most
likely to survive? To this question we are motivated to consider seriously
the following two anomalies in
${R}_{b}$ and ${\alpha }_{s}$.\cite{1a} 
\ba {R}_{b} &=&0.2202\pm
0.0016{\rm\;when\;}{R}_{c} ={R}_{c}^{SM}\nonumber\\
{R}_{b}^{SM} &=& 0.2156\pm 0.003\ea
\ba {\alpha }_{s}({M}_{Z}) &=& 0.126\pm 0.005\pm
0.002{\rm\;using\;LEP\;}{R}_{\ell }{\rm\;only}\nonumber\\
{\alpha }_{s}({M}_{Z}) &=& 0.113\pm
0.005{\rm\;from\;deep\;inelastic\;scattering}\label{ad}\ea

The main reason why these particular
anomalies are intriguing is that the same piece of new physics would account
for both anomalies; namely new physics which slightly enhances the $Zb\overline{b}$
vertex. That is, the shift in this vertex needed to increase
${R}_{b}$ from the standard model value by 2\% would shift the total hadronic
width of the
$Z$ in just such a way so as to reduce the value of
${\alpha }_{s}$ from $R_\ell$ by 0.013. This correlation between the two
anomalies assumes that
$R_c$ is given by the standard model value, since any shift in $R_c$ would
affect both $R_\ell$ and $R_b$. Note that $\Gamma_Z$ and
$\sigma_h^0$ in combination with $R_\ell$ gives a similar result,
${\alpha }_{s}({M}_{Z}) = 0.124\pm 0.004\pm
0.002$, but $\Gamma_Z$ and
$\sigma_h^0$ may be more sensitive to other new physics in addition to
the
$Zb\overline{b}$ vertex. If the main effect of new physics is in the
$Zb\overline{b}$ vertex, then a clear discrepancy should emerge between
${\alpha }_{s}$ from electroweak precision tests and all other measurements of 
${\alpha }_{s}$.

Another point is that if we believe that
the new physics has something to do with the generation of fermion masses, then we
should not be surprised to find this new physics coupling most strongly to the
heaviest family. In fact the correlation between the ${R}_{b}$ and ${\alpha }_{s}$
anomalies became evident \cite{1} through the study of new physics of this sort. In
particular a model \cite{6a} describing a dynamical origin of the $t$-quark mass
contained a new gauge boson $X$ coupling strongly to the third family. This boson
mixed slightly with the $Z$ and thus shifted the $Zb\overline{b}$ vertex. Instead
of starting here by describing a model, we will start from the point of view of
trying to extend the standard model in a simple way, and see where we are led. We
will concentrate on the idea of
$Z$--$X$ mixing, although there are of course other possibilities.\cite{2}

Many authors have
considered the case of an $X$ (often called $Z'$) coupling to all quarks,\cite{3} 
with the goal of accounting for a possible anomaly in ${R}_{c}$ as well as
${R}_{b}$. But we have seen that the correlation between the ${R}_{b}$ and ${\alpha
}_{s}$ anomalies suggests that
${R}_{c}$ should stay at its standard model value. In any case the most recent data
has the possible anomaly in
${R}_{c}$ falling below the $2\sigma$ level.\cite{1a}

We take
$X$ to couple predominantly to the third family, with couplings to the light two
families generated only because of small mass mixing effects. This case has also
been considered in ref. \cite{4}, but in a more conventional
context of elementary scalar fields. There additional Higgs doublets
carrying $X$ charge are postulated to induce the $Z$--$X$ mixing at tree level. It
is found that one additional Higgs coupling to both $t$ and $b$ is not sufficient to
induce mixing with the correct sign. With two additional Higgs---$H_t$ and $H_b$
coupling to
$t$ and $b$ respectively---the desired mixing is possible as long as $H_t$
contributes most of the mixing. Since the $t$ mass violates the $U(1)_X$ gauge
symmetry, it must be generated by the coupling to $H_t$ rather than the
standard model Higgs.

By considering the diagram with the $Z$ coupling to quarks through
an intermediate $X$, via a $Z$--$X$ mass-mixing $\delta {M}^{2}$, the
magnitude of the shift in the $Z$ coupling to quarks can be written as
\be\delta {g}_{Z}=-\delta {M}^{2}{\frac{1}{{M}_{X}^{2}}}{g}_{X}\equiv -\theta
{g}_{X}.\label{a}\ee
To be specific we define the shift in the $Z$ couplings
to be $-\delta {g}_{Z}(\overline{t}{\gamma }_{\mu }{\gamma
}_{5}t+\overline{b}{\gamma }_{\mu }{\gamma }_{5}b)$ and the $X$ coupling to be
$-X^\mu (\overline{t}{\gamma }_{\mu }{\gamma
}_{5}t+\overline{b}{\gamma }_{\mu }{\gamma }_{5}b)$.
(The reason for axial
$X$ couplings to quarks will become clear below.) Since $\delta {M}^{2}$ is the
off-diagonal element of the $2\times 2$ mass-squared matrix, $\theta$ is the
$Z$--$X$ mixing angle which we may assume to be small.

The $Z$--$X$ mixing also
induces a shift in the $Z$ mass, which translates into a contribution to
$\delta \rho $,
\be\delta \rho \approx{\theta }^{2}{\frac{{M}_{X}^{2}}{{M}_{Z}^{2}}}.\label{b}\ee
A possible $Z$--$X$ mixing in the kinetic terms would
contribute a term of the opposite sign,\cite{5,6} but we will assume that this may
be neglected. By inserting
$\theta$ from (\ref{a}) into (\ref{b}) and requiring a large enough $\delta
{g}_{Z}$ to account for
$\delta {R}_{b}$, we have the upper bound
\be{\frac{{M}_{X}}{{g}_{X}}}\lapx 1 {\rm\;TeV}.\label{d}\ee
This is related to the observation made in ref. \cite{4} that ${g}_{X}^{2}/4\pi
\lapx 1$ implies that ${M}_{X}\lapx$ 3--4 TeV.

We further note that $\left\langle{{H}_{t}}\right\rangle$ determines
$\delta {M}^{2}$ and thus \be\delta
{g}_{Z}=-{\frac{e}{sc}}{\left\langle{{H}_{t}}\right\rangle}^{2}
{\frac{{g}_{X}^{2}}{{M}_{X}^{2}}}.\label{e}\ee
Because of the bound (\ref{d}) we have 
\be\left\langle{{H}_{t}}\right\rangle\lapx 50-100{\rm\;GeV}.\label{c}\ee
But since the $t$ mass is generated from $\left\langle{{H}_{t}}\right\rangle$,
(\ref{c}) implies that a large Yukawa coupling is required. By imposing an
upper bound on the size of this Yukawa coupling we now see that both
$\left\langle{{H}_{t}}\right\rangle$ and ${M}_{X}/{g}_{X}$ must come fairly
close to saturating the bounds in (\ref{d}) and (\ref{c}). We have learned two
things; the physics responsible for the $X$ boson mass is characterized by a
TeV, and the
$t$-quark Yukawa coupling is even stronger than in the standard model.

Now let us recall that new fermions must
be introduced to cancel the gauge anomalies involving the $X$. Although new
fermions with nonstandard electroweak charges may be added,\cite{4} a
question is whether a conventional fourth family would suffice. The answer is yes.
If the $X$ has isospin-singlet couplings to the quark doublets $(t,b)$ and
$(t',b')$, then all anomalies are canceled if these two doublets have equal and
opposite {\em vector} $X$ couplings, or equal and opposite {\em axial}
$X$ couplings. $X$ couplings to leptons are not necessary. In fact the two cases
correspond to making different choices for the mass eigenstates. Let us denote by
$Q\equiv (U, D)$ and $Q\equiv (\underline{U},\underline{D})$ the two quark doublets
with equal and opposite vector
$X$ charge. The mass eigenstates
have equal and opposite {\em vector}
$X$ couplings if the mass eigenstates correspond to ${\overline{Q}}_{L}{Q}_{R}$
and
${\overline{\underline{Q}}}_{L}{\underline{Q}}_{R}$. On the other hand the mass
eigenstates have equal and opposite {\em axial} couplings if they correspond to
${\overline{Q}}_{L}{\underline{Q}}_{R}$ and
${\overline{\underline{Q}}}_{L}{Q}_{R}$.

We have noted that ${M}_{X}/{g}_{X}\approx $ 1 TeV, which implies that there is
some physics at a TeV which breaks the $U(1)_X$ gauge symmetry. We now note that if
$t'$ and $b'$ have axial $X$ couplings then their masses do not respect $U(1)_X$.
In this case the existence of these masses is naturally linked to the breakdown of
the $U(1)_X$ at a TeV, implying that the $t'$ and $b'$ masses are of order a
TeV. Given that these new fourth family quarks have conventional weak charges their
masses, if fairly degenerate, would imply appropriate masses for the $W$ and $Z$.
We are being led to consider dynamical electroweak symmetry breaking.

Given this prompting, let us remove all elementary scalar fields. There is then no
tree level contribution to the $Z$--$X$ mass-mixing. There is also little
contribution from
$t'$ and $b'$ loops due to the required degeneracy of the $t'$
and $b'$ masses. That is, the $X$ couplings to $t'$ and $b'$
are the same whereas the axial $Z$ couplings to $t'$ and $b'$ are equal and
opposite, implying that the two contributions in the mass-mixing loop will cancel.
The mixing must then come from the
$t$-loop. It is interesting that the $t$-loop contribution would
vanish if the $X$ had purely vector couplings to the $t$, and so this allows us
to reject the vector coupling possibility.

The shift in the $Z$ coupling from
the $t$-loop is the same as in (\ref{e}), but with
$\left\langle{{H}_{t}}\right\rangle$ replaced by a quantity ${f}_{t}$ determined
by the $t$-loop. ${f}_{t}$ is normalized such that $({f}_{t}/v{)}^{2}$, with
$v\approx $ 240 GeV, gives the fractional contribution of the $t$-loop to
${M}_{Z}^{2}$. The point is that the $t$-loop involves a momentum dependent
$t$ mass function which we may assume is fairly constant up to the scale of new
physics at a TeV, at which point it falls. We thus calculate the loop with a 1
TeV cutoff and find\cite{cdt}
\be{f}_{t}^{2}\approx {\frac{3}{8{\pi
}^{2}}}{m}_{t}^{2}{\rm
ln}\left({{\frac{(1{\rm\; TeV}{)}^{2}}{{m}_{t}^{2}}}}\right)\approx
(60{\rm\;GeV}{)}^{2}.\label{ft}\ee This value is consistent with the constraints we
found before on $\left\langle{{H}_{t}}\right\rangle$. The difference is that
${f}_{t}$ is calculated here, whereas
$\left\langle{{H}_{t}}\right\rangle$ was a free parameter. We conclude that the
$t$-loop produces $Z$--$X$ mixing of the correct magnitude and sign to produce
the desired shift in the $Zb\overline{b}$
vertex. This provides support for our consideration of
dynamical electroweak symmetry breaking.

Leptons of a fourth
family are also appearing in the picture, and their masses must also be large. We
have mentioned that the $X$ boson does not need to couple to leptons to cancel
anomalies, but it is nevertheless easy to motivate such couplings in the context of
quark-lepton unification. It is simplest to expect that the
$U(1)_X$ gauge symmetry commutes with the quark-lepton gauge symmetry present at
some higher scale, in which case the
$X$ boson should couple similarly to quarks and leptons (at least in some basis).
This in turn will shift the
$Z$ couplings to the third (and fourth) family leptons, and the question is
whether such shifts are still allowed by the data. In this connection we note that the
$Z$ couplings to charged leptons is mostly axial. Thus if the shifts occurred mostly
in the vector couplings then the strongly constrained leptonic partial
decay widths of the
$Z$ would be little affected.

Anomalies
will cancel within the lepton sector if the two families of leptons $({\nu
}_{L},{\tau }_{L},{\tau }_{R};$ ${\nu
}_{L}^{\prime },{\nu }_{L}^{\prime },{\tau }_{R}^{\prime })$  have $X$ charges
$(+,+,+;-,-,-)$ or $(+,+,-;-,-,+)$. The difference again is related to choice of
mass eigenstates. As for the quarks we may define the fields $({E}_{L},{E}_{R})$ and
$({\underline{E}}_{L},{\underline{E}}_{R})$ to have equal and opposite vector
$X$ charge. Unlike for quarks these fields must correspond to the mass
eigenstates, so that we have vector
$X$ couplings to the
$\tau $ ($E$) and ${\tau }^{\prime }$ (${\underline E}$), and thus shifts mainly in
the vector rather than the axial $Z$ coupling to $\tau$. We emphasize that the
underlying strong dynamics at a TeV is responsible for the choice of mass
eigenstates, since it determines the fields corresponding to the fourth family
quarks and leptons.

We have thus motivated the following $X$ boson coupling to the
third family,
\be{J}_{\mu }^{X}=\overline{t}({L}_{\mu }-{R}_{\mu })t+\overline{b}({L}_{\mu
}-{R}_{\mu })b+\overline{\tau }({L}_{\mu }+{R}_{\mu })\tau +{\overline{\nu
}}_{\tau }{L}_{\mu }{\nu }_{\tau }\ee
with $L_\mu,R_\mu \equiv \gamma_\mu(1\mp\gamma_5)/2$. The most striking prediction
is that if the ${R}_{b}$ anomaly is to be explained as we have described, then the
$\tau$ asymmetry parameter ${{\cal A}}_{\tau }$ would become $\approx$ 20\% higher
than in the standard model.\cite{1} If we assume that ${{\cal A}}_{e}$ is given by
the standard model (an assumption in good agreement with LEP data), then there
are two independent measurements of ${{\cal A}}_{\tau }$. The forward-backward
asymmetries and the mean tau polarization give \cite{1a}
\begin{eqnarray}{A}_{FB}^{0,\tau }&\Rightarrow & {{\cal A}}_{\tau }/{{\cal A}}_{\tau
}^{SM}=1.28\pm 0.14,\\ 
{\cal P}&\Rightarrow & {{\cal A}}_{\tau }/{{\cal A}}_{\tau }^{SM}=0.96\pm
0.05.\end{eqnarray} When combined these results are in perfect agreement with the
standard model, but the discrepancy between the results may suggest that it is too
early to completely rule out new physics in the $Z$ coupling to $\tau$.

We note
that predicted shifts in other quantities due to $Z$--$X$ mixing, $+0.2$\%,
$-1.5$\%, and $-0.5$\% in
${\Gamma }_{\tau}$, ${\Gamma }_{\nu_\tau}$, and ${{\cal A}}_{b}$
respectively,\cite{1} are all compatible with current measurements. Additional small
corrections to ${\Gamma }_{\tau}$ from vertex corrections and to $\delta
\rho$ from two-loop graphs are discussed in \cite{6c} and \cite{7} respectively. In
the event that the
$X$ couples only to quarks, then only the shifts in
$\Gamma_b$ and
${{\cal A}}_{b}$ remain. We also note \cite{1,2} that flavor changing neutral currents
induced by nonuniversal $Z$ couplings are acceptable as long as most mass mixing
occurs in the up-quark sector.

\section{Neutrino Mass and Quark Mass}

We have seen that the $Z$--$X$
mixing can be induced by a Higgs as long as that Higgs has a large Yukawa coupling to
the $t$ quark. But then we saw that the $t$-loop would suffice by itself, so
that we could remove the Higgs. We now note that a similar situation occurs when we
consider the massive fourth-family neutrino. We treat the
case where all right-handed neutrinos are absent from the effective theory at a TeV,
since we note that the large Majorana masses for right-handed neutrinos are allowed
by the $U(1)_X$ gauge symmetry. We must then consider the electroweak corrections
induced by the fourth-family left-handed neutrino. Other analyses\cite{6b} either
have Dirac neutrino masses or right-handed neutrinos involved in a see-saw mechanism.
But in the dynamical symmetry breaking context it appears that right-handed neutrinos
much more massive than a TeV will more naturally completely decouple.

The fourth-family left-handed neutrino
${\nu }_{L}$ must have a Majorana mass greater than
${M}_{Z}/2$. If this mass were to come from a vacuum expectation value of a
$SU(2{)}_{L}$-triplet scalar field
$\left\langle{{H}_{M}}\right\rangle$, then we have a tree level contribution
to $\alpha T\approx
-(\left\langle{{H}_{M}}\right\rangle/125{\rm\;GeV}{)}^{2}$. This puts a severe upper
bound on $\left\langle{{H}_{M}}\right\rangle$, which in turn implies a very
large
${\nu }_{L}$ Yukawa coupling. We therefore are in the same situation as
before, leading us again to remove scalar fields and consider the dynamical
generation of mass. Of interest now is the neutrino-loop contribution to $T$.

We have considered \cite{6} the contributions to
$S$, $T$, and $U$ from the fourth family leptons $({\nu }_{L},{\tau }^{\prime
})=(N,E)$ (we omit the underlines) for a range of masses ${m}_{N}$ and ${m}_{E}$. The
result for
$T$ depends on the effective cutoff in the neutrino loop (similar to
(\ref{ft})); this cutoff is supplied by the momentum dependence of the
dynamical mass, and we have used $\Lambda = 1.5m_N$ and $\Lambda = 2m_N$.
From Figs. (1) and (2) we see ranges of masses for which the new
contributions to
$T$ are negative with reasonable size, while the new contributions to $S$ and $U$ are
simultaneously small. We therefore have what appears to be a natural source of
negative
$T$ within the dynamical symmetry breaking context. This may be useful, given the fact
that a dynamically generated $t$-quark mass is typically accompanied by
positive contributions to $T$.

Let us consider quark masses further. ${t}^{\prime }$ and ${b}^{\prime }$
have received TeV masses via dyanamics associated with the breakdown of $U(1)_X$. In
the absence of scalars there must be four-fermion operators which will feed mass down
to the other quarks, and the largest such operator will be the one which provides the
$t$ mass. We recall our previous notation for the two quark doublets having equal and
opposite vector $X$ charge:
$Q\equiv (U,D)$ and $Q\equiv (\underline{U},\underline{D})$. The fourth family
quarks have the form ${\overline{\underline{Q}}}_{L}{Q}_{R}$, which breaks $U(1)_X$,
and we need an operator to feed this down to
${\overline{Q}}_{L}{\underline{Q}}_{R}$. A suitable operator which will feed mass
from the $b'$ to the $t$ is
$({\overline{Q}}_{L}{D}_{R})\epsilon
({\overline{\underline{Q}}}_{L}{\underline{U}}_{R})$. (The antisymmetric
$\epsilon$ has $SU(2)_L$ indices.) Note that this operator is composed of two
Lorentz scalars which are also singlets under
$U(1)_X$. This operator is thus of a form which could be expected to be enhanced by
strong
$U(1)_X$ interactions. If these interactions are of the walking-coupling type,
then there can be significant anomalous scaling enhancement of this operator relative
to other operators.

This operator has a partner,
$({\overline{Q}}_{L}{U}_{R})\epsilon
({\overline{\underline{Q}}}_{L}{\underline{D}}_{R})$. This operator will feed
mass from the $t'$ to the $b$, and it must thus be suppressed relative to the
previous operator. The attractive feature is that the isospin breaking implied
by the different sizes of these
operators does not feed into the TeV quark masses or $T$ in a direct way. In
fact four insertions of these operators are needed to produce a contribution
to $T$. Other isospin breaking operators which could feed more directly into
$T$ are not so enhanced by the anomalous scaling effects. We thus have an
example \cite{6a} of electroweak breaking physics being protected to some extent from
the isospin breaking physics, which is feeding down from higher scales in
four-fermion operators.

We may bring in the two light families by embedding
$U(1)_X$ into a larger gauge symmetry at a higher scale $\Lambda$ (say
100--1000 TeV), which connects the light families to heavy families. We may
then write down various four-fermion operators which can arise at the scale
$\Lambda$. It may be shown that such operators are sufficient to produce nontrivial
mass mixing and a potentially realistic mass matrix.\cite{7} We find that the
existence of these operators may also be associated with the breakdown of the
$U(1)_X$ gauge symmetry.

Hierarchies develop in the quark mass matrices for three reasons. 1) The
various operators have different numbers of fields to which the
$U(1)_X$ couples, and thus are enhanced by varying amounts due to anomalous
scaling induced by the $U(1)_X$. 2) Some entries in the mass matrices
receive mass fed down from a
$t$ or ${\tau }^{\prime }$ rather than the heavier $t'$ or $b'$. 3) Some
entries only receive contributions from loops involving more than one
4-fermion operator. We note from \cite{7} that contributions feeding down
from the ${\tau }^{\prime }$ are important for obtaining realistic
quark mass matrices, and that these contributions only arise for the choice
of mass eigenstates for $\tau$ and ${\tau
}^{\prime }$ which we have already motivated.

\section*{Acknowledgments} This research was supported in part by the Natural Sciences and Engineering
Research Council of Canada.

\newpage
\begin{figure}
\begin{center}
\leavevmode
\epsffile{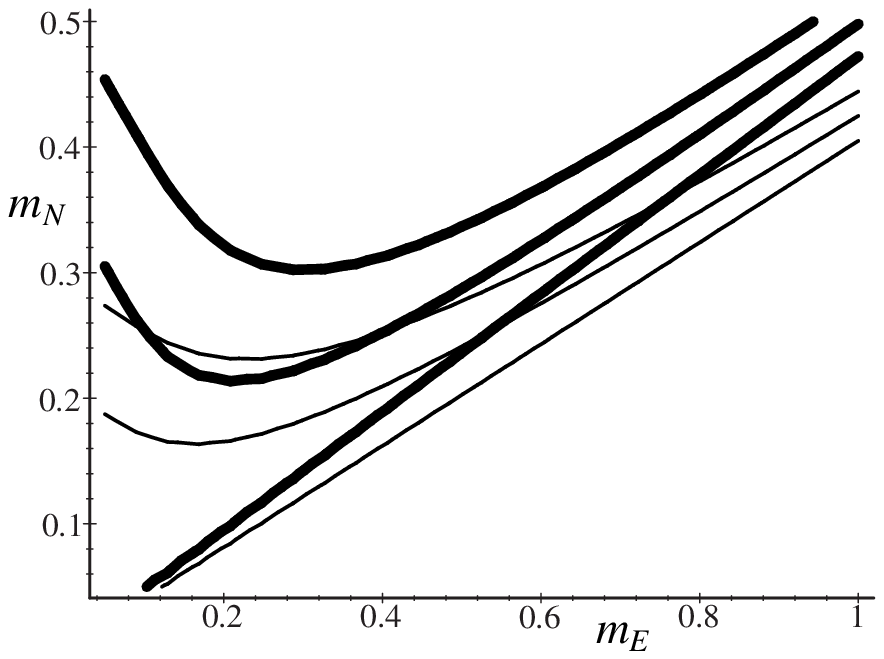}
\end{center}
\caption{Lines of constant $T$ as a function of the $N$ and $E$ masses in TeV. Thick
and thin lines are for $\Lambda = 1.5m_N$ and $\Lambda = 2m_N$ respectively. In
each case, from top to bottom, $T=-2,-1,0$.}
\end{figure}
\begin{figure}
\begin{center}
\leavevmode
\epsffile{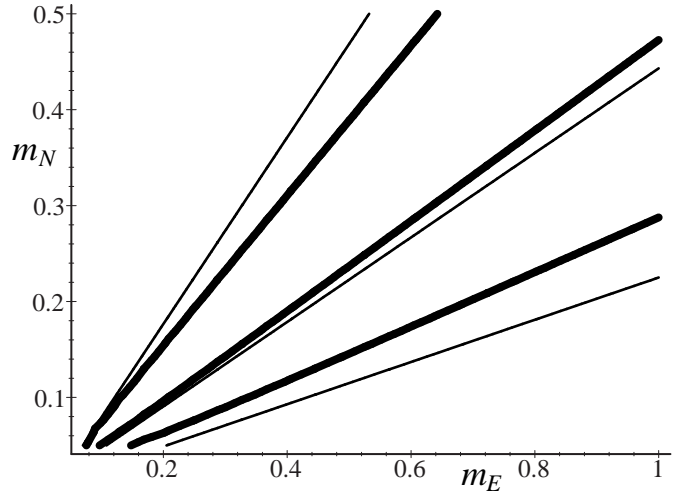}
\end{center}
\caption{Thick and thin lines are lines of constant $S$ and $U$ respectively as a
function of the $N$ and $E$ masses in TeV. From top to bottom in each case
$S=1/6\pi,0,-1/6\pi$ and $U=-1/12\pi,0,1/6\pi$.}
\end{figure}
\end{document}